\documentclass{article}
\usepackage{graphicx} 
\usepackage{authblk}
\usepackage{cite}

\begin{document}

\title{Comment on `Painlev\'e Integrability and Multi-wave pattern for (2+1) dimensional Long wave-Short wave Resonance Interaction system'}
\author[$\dagger$]{R. Radha}
\author[$\star$]{C. Senthil Kumar}
\affil[$\dagger$]{Centre for Nonlinear Science (CeNSc), Post-Graduate and Research Department of Physics, Government College for Women(Autonomous), Kumbakonam, 612 001, Tamil Nadu, India}
\affil[$\star$]{Department of Physics, Vinayaka Mission's Kirupananda Variyar Engineering College, Vinayaka Mission's Research Foundation (Deemed to be University), NH-47, Sankari Main Road, Periaseeragapadi, Salem 636 308, Tamil Nadu, India}
\maketitle
\begin{abstract}
In the paper by Sivatharani et al [1], the authors make a tall claim about the integrability of a 2 component (2+1) dimensional Long wave Short wave Resonance Interaction (2C(2+1)LSRI) equation with mixed sign  which was already claimed to be non integrable and hence known not to satisfy Painlev\'e [2] property which the authors show to pass Painlev\'e test.  We have categorically shown how the system does not pass Painlev\'e test and hence non-integrable reinforcing the claim made by Maruno et al [2].  The authors claim to derive the solutions of 2C(2+1)LSRI equation which ironically do not satisfy the equation.  To top it all, the    authors claim to have generated lumps and  dromions which again defy their very own definition.
\end{abstract}

{\textbf{Keywords:}}{\textit{Painlev\'e property, integrability, dromions, lumps}



\section{Introduction}
The 2C(2+1) LSRI equation is given by
\begin{eqnarray}
    i(A_t+A_y)-A_{xx}+LA=0, \\
    i(B_t-B_y)-B_{xx}+LB=0, \\
    L_t=2({|A|^2_x}+{|B|^2_x}).
\end{eqnarray}
The above equation was derived by Maruno et al \cite{Maruno} in a two layer fluid using reductive perturbation method and they postulated that it belongs to the class of nonintegrable equations. The 2C(2+1)LSRI system does not possess Painlev\'e property  as claimed  by the authors \cite{Sivatharani}. 
To substantiate our claim, we bring out the non-Painlev\'e nature of 2C(2+1)LSRI equation.

\section {\bf Singularity Structure Analysis and 
Non-Painlev\'e Nature}
Expanding the field variable and its conjugate in the form of a Laurent series in terms of a non-characteristic singular manifold and proceeding in accordance with \cite {Weiss},  the authors \cite{Sivatharani} obtain the resonance values as 
\begin{eqnarray}
    j=-1,0,0,0,2,3,3,3,4. \;\;\;\;\;\;\;\;\;(17) \nonumber
\end{eqnarray}
The resonance at $j=-1$ can be associated with  the arbitrariness of the manifold $f(x,y,t)=0$ while   the resonances at $j=0,0,0$ are associated with the following leading order behaviour of the solutions
\begin{eqnarray}
  \alpha_0\beta_0+\gamma_0\delta_0 = f_x f_t. \;\;\;\;\;\;\;(45) \nonumber
\end{eqnarray}
Proceeding  further to establish the existence of arbitrary function at    resonance $r=2$, the authors \cite{Sivatharani}  obtain the following equations  
\begin{eqnarray}
   & & i (\alpha_{0t}+\alpha_{0y})-\alpha_{0xx}+L_0\alpha_2+L_1\alpha_1+L_2\alpha_0 = 0, \;\;\;\;\;\;\;(29) \nonumber\\
   & &  -i (\beta_{0t}+\beta_{0y})-\beta_{0xx}+L_0\beta_2+L_1\beta_1+L_2\alpha_0=0, \;\;\;\;\;\;(30) \nonumber\\
   & &  i (\gamma_{0t}-\gamma_{0y})-\gamma_{0xx}+L_0\gamma_2+L_1\gamma_1+L_2\gamma_0=0, \;\;\;\;\;\;\;\;\;\; (31) \nonumber\\
   & &  -i (\delta_{0t}-\delta_{0y})-\delta_{0xx}+L_0\delta_2+L_1\delta_1+L_2\delta_0=0, \;\;\;\;\;\;\;\;\;(32) \nonumber\\
   & & L_{1t}=2[\alpha_{0x}\beta_1+\alpha_0\beta_{1x}+\alpha_{1x}\beta_0+\alpha_1\beta_{0x}] \nonumber\\
    & & \;\;\;\;\;\;\;\;+2[\gamma_{0x}\delta_1+\gamma_0\delta_{1x}+\gamma_{1x}\delta_0+\gamma_1\delta_{0x}]. \;\;\;\;\;\;\;\;\;\;\;\;\;\;\;\;\;\;\;\; (33) \nonumber
\end{eqnarray}
When we substitute the leading order results ($\alpha_0, \beta_0,\gamma_0, \delta_0$) and $\alpha_1, \beta_1,\gamma_1, \delta_1, L_1$ given by the authors \cite {Sivatharani} in equation (33), we observe that equation (33) becomes incompatible. This means that 2C(2+1)LSRI mixed case system does not satisfy the Painlev\'e property \cite{Weiss} and this observation reflects the non-integrable nature of the system which is consistent with the observation of Maruno et al \cite{Maruno}. Hence, the claim of the authors\cite{Sivatharani} stands invalid.

\section{Validation of Solutions}
\begin{enumerate}
    \item The solutions derived by the authors employing Truncated Painlev\'e approach \cite{Radha05}  are given in equations (85)-(88).  Ironically, these solutions do not satisfy the 2C(2+1) LSRI system.  In particular, the solutions satisfy one part of 2C(2+1)LSRI equation alone i.e  equation (1) in \cite{Sivatharani} .  Substituting the solutions in the remaining  parts of the system i.e. namely in equations (2) and (3)  \cite{Sivatharani} renders them invalid as they do not satisfy equations (2) and (3). Hence, the solutions derived by the authors stand invalid.
    \item A closer look at the solutions indicate that they all (rogue waves, lumps and dromions) undergo change in amplitude with time. This perception totally goes against their very definition. While the rogue waves which are driven by rational functions  are known to be unstable with their amplitude varying with time, lumps (multi lumps) do not interact with each other. On the other hand,  dromions (exponentially localized solutions) have been known to undergo both elastic \cite{Radha05} and inelastic \cite{Radha22} collision and have never been proven  to be unstable. In the light of the above observation,  the lump solution given by the authors (Figure 3) where the amplitude grows  with time (oscillates with time)  is totally unacceptable and incorrect.  Hence, this is not a lump at all.
    \item In Figure 4 again, the amplitude of the dromion varies as time progresses.  But, as per the definition, for isospectral (2+1) dimensional nonlinear partial differential equations (pdes),   the amplitude of a dromion remains constant unless otherwise it undergoes collision with another dromion. The amplitude of dromions can vary with time only for non-isospectral nonlinear pdes while 2C(2+1)LSRI equation is isospectral in nature. Again, this defies the very definition of a dromion. 
    \item In  Figure 5 containing  the two-dromion solution plotted by the authors \cite{Sivatharani}, the amplitude of the two-dromion changes as time evolves from t=-55 to +55. The authors claim to have found dromions with unstable amplitudes, an observation  which can never arise. The above fact is indeed a proof of the erroneous nature of the solutions.  
\end{enumerate}

\section{Mathematical Errors}
The following errors were observed.
\begin{enumerate}
 \item Very first equation is incorrect.  Equation (3) in \cite{Sivatharani} should be $L_t$ instead of $L_y$.
    \item In equation (4) in \cite{Sivatharani}, the sign of the third term $\alpha_{xx}$ should be -ve instead of +ve.
     \item In equation (5) in \cite{Sivatharani}, the sign of the third term $\beta_{xx}$ should be -ve instead of +ve.
      \item In equation (6) in \cite{Sivatharani}, the sign of the third term $\gamma_{xx}$ should be -ve instead of +ve.
       \item In equation (7) in \cite{Sivatharani}, the sign of the third term $\delta_{xx}$ should be -ve instead of +ve.
       \item In equation (15) in \cite{Sivatharani}, in the second and third terms, there is a sudden change in notation, when we observe the flow of the paper through equations (9) \& (10) in \cite{Sivatharani}.  The fifth part of Equation (9) in \cite{Sivatharani} reads as $L = L_0f^m$.  The second part of the equation (10) in \cite{Sivatharani} reads as $m=-2$.  But in equation (15) in \cite{Sivatharani}, there is a change in notation from $m$ to $w$. Equation (15) in \cite{Sivatharani} reads as $L = L_0 f^w +.....+ L_j f^{j+w}+....,$. Instead, it should be $L = L_0 f^m +.....+ L_j f^{j+m}+....,$.
    \item In equation (24) in \cite{Sivatharani}, the fourth term should be $-i\alpha_0 f_y$ instead of $-i\alpha_0 f i_y$.
     \item In equation (25) in \cite{Sivatharani}, the fourth term should be $i\beta_0 f_y$ instead of $i\beta_0 f i_y$.
      \item In equation (26) in \cite{Sivatharani}, the fourth term should be $i\gamma_0 f_y$ instead of $i\gamma_0 f i_y$.
       \item In equation (27) in \cite{Sivatharani}, the fourth term should be $-i\delta_0 f_y$ instead of $-i\delta_0 f i_y$.
       \item In equation (30) in \cite{Sivatharani}, the second term should be $\beta_{0y}$ instead of $q_{0y}$.
\end{enumerate}

\section{Conclusion}
To conclude, the 2 Component (2+1) dimensional Long wave-Short wave Resonance Interaction (2C(2+1)LSRI) equation with mixed sign case does not pass Painlev\'e test as claimed by the authors and hence  non-integrable.  The lumps and dromions generated defy their characteristics and do not satisfy  the equation.  Lumps, dromions and two dromion solutions plotted are also incorrect.

\section{Declaration}
RR wishes to thank Council of Scientific and Industrial Research (CSIR), Government of India, for financial support under grant No 13(1456)/EMR-II.

\section{Data Availability}
The datasets analysed during the current study are not publicly available due to the huge size of calculations, but are available from the corresponding author on reasonable request.


\begin{thebibliography}{10}
\bibitem{Sivatharani} B. Sivatharani, K. Subramanian, A. Sekar, P. Shanmuga Sundaram, Painlev\'e Integrability and Multi-wave pattern for (2+1) dimensional long wave-Short wave resonance interaction system,  Nonlinear Dyn., {\bf 109}, 1935–1946 (2022).
\bibitem{Maruno} K. Maruno, Y. Ohta, M. Oikawa, Note on the Two-component analogue of Two-dimensional Long wave-Short wave Resonance Interaction Equation, Glasgow. Math. J. {\bf 51}A  129-135 (2009). 
\bibitem{Weiss} J. Weiss, M. Tabor, G. Carnevale, The Painlevé property for partial differential equations. J. Math. Phys., {\bf 24} 522-526 (1983).
\bibitem{Radha05} R. Radha, C. Senthil Kumar, M. Lakshmanan, X. Y. Tang and S. Y. Lou, Periodic and Localized Solutions of Long wave-Short wave Resonance Interaction Equation, J. Phys. A: Math. Gen. {\bf 38}, 9649-63 (2005).
\bibitem{Radha22} R. Radha, C. Senthil Kumar, Localized Excitations and their collisional dynamics in (2+1) dimensional Broer-Kaup-Kupershmidt equation, Rom. Rep. Phys. {\bf 74}, 104 (2022).
\end{thebibliography}
\end{document}